\documentclass[nofootinbib,prd,aps,superscriptaddress,notitlepage,reprint,preprintnumbers,showpacs,showkeys]{revtex4-1}
\usepackage{latexsym,graphicx,amssymb,amsmath,mathrsfs} 
\usepackage[utf8]{inputenc}
\usepackage{latexsym,graphicx,amssymb,amsmath}
\usepackage{setspace,bm}

\DeclareSymbolFont{bbold}{U}{bbold}{m}{n}
\DeclareSymbolFontAlphabet{\mathbbold}{bbold}

\newcommand{\be}{\begin{equation}}      
\newcommand{\ee}{\end{equation}}      
\newcommand{\bea}{\begin{eqnarray}}      
\newcommand{\eea}{\end{eqnarray}}    
     
\newcommand{\rt}[1]{{}}

\newcommand{\Tr}{\,\textrm{Tr}\,}

\newcommand{\rhs}{\,\textrm{rhs}\,} 
\newcommand{\lhs}{\,\textrm{lhs}\,}

\makeatletter
\renewcommand\appendix{\par
\setcounter{section}{0}%
\setcounter{subsection}{0}%
\gdef\thesection{\appendixname\space\@Alph\c@section}}

\long\def\unmarkedfootnote#1{{\long\def\@makefntext##1{##1}\footnotetext{#1}}}
\makeatother

\begin{document} 

\title{Flows of multicomponent scalar models with $U(1)$ gauge symmetry} 

\affiliation{Department of Physics, Keio University, Yokohama 223-8521, Japan}
\affiliation{Institute of Physics, E{\"o}tv{\"o}s University, Budapest 1117, Hungary}
\affiliation{RIKEN Interdisciplintary Theoretical and Mathematical Sciences Program (iTHEMS), Wako 351-0198, Japan}
\affiliation{RIKEN Nishina Center, Wako 351-0198, Japan}
\author{G. Fej\H{o}s}
\email{fejos@keio.jp}
\affiliation{Department of Physics, Keio University, Yokohama 223-8521, Japan}
\affiliation{Institute of Physics, E{\"o}tv{\"o}s University, Budapest 1117, Hungary}
\affiliation{RIKEN Nishina Center, Wako 351-0198, Japan}
\author{T. Hatsuda}
\email{thatsuda@riken.jp}
\affiliation{RIKEN Interdisciplintary Theoretical and Mathematical Sciences Program (iTHEMS), Wako 351-0198, Japan}
\affiliation{RIKEN Nishina Center, Wako 351-0198, Japan}

\begin{abstract}
{We investigate the renormalization group flows of multicomponent scalar theories with $U(1)$ gauge symmetry using the functional renormalization group method. The scalar sector is built up from traces of matrix fields that belong to simple, compact Lie algebras. We find that in these theories the local potential approximation (LPA) is not a one-loop closed truncation in general, even at zero gauge coupling. If, however, we add a $U(1)$ factor to the Lie algebra structure, then the LPA always becomes one-loop closed. In accordance with our earlier findings, fluctuations introduce anomalous, regulator dependent gauge contributions, which are only consistent with the flow equation for a given set of gauge fixing parameters. We establish connections between regularization procedures in the standard covariant and the $R_\xi$ gauges arguing that one is not tied by introducing regulators at the level of the functional integral, and it is allowed to switch between schemes at different levels of the calculations. We calculate $\beta$ functions, classify fixed points, and clarify compatibility of the flow equation and the Ward-Takahashi identity between the scalar wave function renormalization and the charge rescaling factor.}
\end{abstract}

\maketitle

\section{Introduction}

Modern implementation of the idea of the Wilsonian renormalization group (RG), the functional or exact RG, has had great success in the past for field theories with global, linearly realized symmetries \cite{wetterich93,morris94}. One of the key ingredients is that these types of symmetries can be exactly implemented in the functional integral representation of the scale dependent quantum effective action. Unfortunately, local gauge symmetries and nonlinear symmetries require a much more careful treatment as masslike deformations, such as the regulator term in the functional RG (FRG) formalism explicitly break them.

In gauge theories the Ward-Takahashi (or Slavnov-Taylor) identities of gauge symmetry remain, but by construction, they get corrected by terms coming from the infrared (IR) regulator \cite{ellwanger94,bonini94,dattanasio96,igarashi00,freire01,igarashi00b,igarashi07}. These so-called modified Ward-Takahashi identities (mWTIs) have been shown to be compatible with the scale evolution equation in the sense that if they are satisfied at any scale, then they are satisfied at all scales, given that the effective action obeys the flow equation. This statement is sometimes argued to be violated by approximate solutions \cite{gies12}, but in our earlier work we found compatibility \cite{fejos17}, and in this paper, we also aim to provide further evidence that in the local potential approximation the flow equation and the mWTIs lead to the same scale dependence of the couplings. Once the regulator is removed, the mWTIs reduce to the standard WTIs; therefore, one expects gauge symmetric results in the infrared. In practice, the main problem with this is that if one is to seek for scaling solutions of the flow equation, the IR regulator is never fully removed (otherwise, the scaling could not be seen whatsoever), and the aforementioned anomalous terms in the Ward-Takahashi identities can indeed have significance. They can lead to the absense of IR fixed points, or signal fake solutions that are not supposed to be found in the continuum theory. 

The unsettling nature of gauge symmetry violation have been tackled by several methods. The background field method, where gauge invariance is maintained under background field transformations, has been a popular scheme \cite{reuter94,reuter94b,bergerhoff96,bergerhoff96b,gies02}; nevertheless, quantum gauge invariance is still encoded in modified identities \cite{safari16,safari17}, being treated only approximately. Manifestly gauge invariant flow equations have been proposed without the Fadeev-Popov method \cite{morris01,arnone02,arnone03,arnone07} and also for the geometric effective action, via the Vilkovisky-DeWitt framework  \cite{branchina03}. By suitable definition of macroscopic gauge fields, a new version of a gauge-invariant flow equation has also appeared \cite{wetterich18}, which resembles to the background field method in a specific gauge. Recent attempts showing that gauge, or in principle Becchi-Rouet-Stora-Tyutin (BRST) symmetry is not necessarily broken by the presence of a cutoff can be found in \cite{asnafi18,igarashi19}.  Despite the conceptual successes of implementing gauge invariance into the RG flows, from a technical point of view, and thus considering practical computations, still the standard quantization proves to be the most easily accessible method.

In this study, we also choose to proceed this way and deal with gauge symmetry violation through a gauge fixing that is tailored to the approximation we use. Our aim is to extend earlier results on the $U(1)$ gauge theory with $N$ complex scalars \cite{fejos17}. On the one hand, we are interested in a family of theories, where the scalars ($\Phi$) belong to the fundamental representation of an unspecified Lie algebra in a way that allows us to have two independent quartic couplings in the classical renormalizable potential, as operators $\sim \!|\Tr[\Phi^\dagger \Phi]|^2$ and $\sim \!\Tr [\Phi^\dagger \Phi \Phi^\dagger \Phi]$. Scalar sectors of this type are present, e.g., in meson models, or in the effective theory of color superconductivity. On the other hand, we wish to perform our investigations in the usual covariant gauge, i.e., $\sim\!(\partial_i {\cal A}_i)^2/2\xi$ (here ${\cal A}_i$ is the gauge field), rather than in the $R_\xi$ gauge,  which we used in a previous study \cite{fejos17}. Even though the latter was very convenient from several computational points of view, it did not respect the symmetry generated by the interchange between the real and imaginary parts of the scalars. This prevented us from performing a complete check of compatibility between the flow equation and the regulator modified Ward-Takahashi identities; see details in \cite{fejos17}. In this paper, we wish to rederive and extend our earlier results, but now in the covariant gauge, show the aforementioned compatibility and draw some new conclusions regarding the interplay between regularization schemes and gauge fixing terms. We believe that these contributions help facilitate a deeper understanding of the application of the FRG method in gauge theories and opens up new approximations for the future. As a general outcome of our method, all $\beta$ functions of the couplings will be calculated analytically, which makes it possible to find and classify known and new fixed points in the system. It is particularly interesting to investigate what type of charged fixed points (i.e. with nonzero gauge coupling) can appear.

The paper is organized as follows. After discussing the basics of the method in Sec. II, Sec. III is devoted to investigating what scalar theories are one-loop closed from a RG flow point of view, without coupling them to any gauge field. This will turn out to be nontrivial regarding the structure of the underlying Lie algebra. Once scalar theories of interest are specified, in Sec. IV, we turn on the gauge coupling and investigate under what circumstances consistency between the RG flow and gauge symmetry survives, and reveal how to construct equivalent regulators in the covariant and $R_\xi$ gauges. We will also calculate the $\beta$ functions of all couplings, classify the existing fixed points, and investigate the connection between the flow equation and the Ward-Takahashi identity of the scalar wave function renormalization and the charge rescaling factor. The reader finds the summary and outlook in Sec. V.

\section{Basics}

Euclidean Lagrangians of the family of theories that are to be investigated in this paper take the following form:
\bea
\label{Eq:basicL}
L &=& \frac{{\cal A}_i}{2}\left(-\partial^2 \delta_{ij} + (1-\xi^{-1})\partial_i \partial_j\right){\cal A}_j + \Tr (D_i \Phi^\dagger D_i \Phi) \nonumber\\
&+&\mu^2\Tr(\Phi^\dagger \Phi)+g_1 |\Tr(\Phi^\dagger \Phi)|^2+g_2 \Tr (\Phi^\dagger \Phi \Phi^\dagger \Phi), \nonumber\\
\eea
where ${\cal A}_i$ is a $U(1)$ gauge field, $\Phi=(s^a+i\pi^a)T_a$, $T_a$ are generators of a $d$ dimensional Lie algebra that generates a compact Lie group, i.e., they are Hermitian [$T_a=(T_a)^\dagger$], $D_i=\partial_i + i e {\cal A}_i$ is the covariant derivatve, and we have also added the usual covariant gauge fixing term with a $\xi$ gauge fixing parameter. The generators are normalized as $\Tr (T_a T_b) = \delta_{ab}/2$. We are interested in that, what circumstances (\ref{Eq:basicL}) are compatible with the renormalization group flow. That is to say, if one considers the scale dependent effective action $\Gamma_k$ (which contains all fluctuations beyond scale $k$), is it always true that if one starts with renormalizable operators at the UV scale, $\Gamma_k$ preserves that structure and does not lead to noncancelable divergences? This question is usually answered via the help of symmetries: for linearly realized global symmetries of the classical action, it is quite trivial to show that the effective action respects those symmetries, and thus, no additional terms are generated in the effective action that could lead to divergences in a continuum theory. For nonlinearly realized symmetries (such as non-Abelian gauge symmetry), this is less trivial, but they are also of textbook examples \cite{zinnjustin}.

Here, we pose the question differently: without specifying any symmetry of the theory, what is the requirement for the underlying Lie algebra structure that leads to renormalizable theories? Furthermore, how is this affected by the $U(1)$ gauge field? Right from the beginning, we wish to be clear on that we are not aiming to provide any rigorous mathematical proof to either of these questions, but we wish to investigate if the local potential approximation, that is, $\Gamma_k = \int L_k$,
\bea
\label{Eq:basicG}
L_k &=&  Z_{A,k}\frac{{\cal A}_i}{2}\left(-\partial^2 \delta_{ij} + (1-\xi_k^{-1})\partial_i\partial_j\right){\cal A}_j \nonumber\\
&+& Z_k \Tr (\hat{D}_i \Phi^\dagger \hat{D}_i \Phi)+V_k,\nonumber\\
V_k&=&Z_k\mu_k^2 \Tr(\Phi^\dagger \Phi)  \nonumber\\
&+&Z_k^2g_{1,k} |\Tr(\Phi^\dagger \Phi)|^2+Z^2_kg_{2,k} \Tr (\Phi^\dagger \Phi \Phi^\dagger \Phi)
\eea
is one-loop closed; i.e., the structure of the classical action is preserved by the flow equation. Here, $\hat{D}_i=\partial_i +i e {\cal A}_i Z_{e,k}/Z_k$, and (\ref{Eq:basicG}) is obtained from (\ref{Eq:basicL}) via the following rescalings: $\Phi \rightarrow Z_k^{1/2} \Phi$, ${\cal A}_i \rightarrow Z_{A,k}^{1/2} {\cal A}_i$, $e \rightarrow eZ_{e,k}/Z_kZ_{A,k}^{1/2}$. 

The evolution of the $\Gamma_k$ scale dependent effective action is given by \cite{wetterich93}
\bea
\label{Eq:flow1}
k\partial_k \Gamma_k = \frac 12 k\tilde{\partial}_k \Tr \log \left(\Gamma_k^{(2)}+{\cal R}_k\right),
\eea
where $\Gamma_k^{(2)}$ is the second functional derivative matrix of $\Gamma_k$ with respect to all field variables, and ${\cal R}_k$ is a regulator function (it is also a matrix in the inner space of fields), which is meant to suppress fluctuations with momenta $|q| \lesssim k$. In (\ref{Eq:flow1}), $\tilde{\partial}_k$ acts only on the regulator and throughout the paper, unless stated otherwise, we use the optimized version \cite{litim01}, i.e., in Fourier space, ${\cal R}_k(q,p)=(2\pi)^{\cal D}{\cal R}_k(q)\delta(p+q)$, ${\cal R}_k(q)=\hat{{\cal Z}}_kR_k(q)$, where $R_k(q)=(k^2-q^2)\Theta(k^2-q^2)$, and $\hat{{\cal Z}}_k$ is the coefficient matrix of the $q^2$ terms in the diagonal entries of $\Gamma_k^{(2)}$. $\Theta(x)$ is the step function, and ${\cal D}$ is the spacetime dimension. In (\ref{Eq:flow1}), the $\Tr$ operation has to be taken both in the functional and in the matrix sense. 

It is useful to reformulate (\ref{Eq:flow1}) in the following way. We separate in $\Gamma_k^{(2)}$ the Gaussian part $\Gamma_k^{(2)0}$ from the interactions as
\bea
\label{Eq:Gamma2}
\Gamma_k^{(2)}=\Gamma_k^{(2)0}+U_k'',
\eea
where the primes indicate all field differentiations, and this relation also {\it defines} $U_k$. Then we introduce the notation $\Gamma_{k,R}^{(2)0}\equiv \Gamma_{k}^{(2)0} + {\cal R}_k$ to reformulate the flow equation as
\bea
\label{Eq:flow2}
k\partial_k \Gamma_k &=& k\tilde{\partial}_k \Tr \log \Gamma_{k,R}^{(2)0} \nonumber\\
&+&\frac12 k\tilde{\partial}_k \Tr \log \left(1+ (\Gamma_{k,R}^{(2)0})^{-1}U_k''\right), 
\eea
where the first term can be discarded as it is an irrelevant constant. The form (\ref{Eq:flow2}) is convenient, because $\Gamma_{k,R}^{(2)0}$ is easily invertable even for the case of inhomogeneous field configurations, and projecting (\ref{Eq:flow2}) onto various operators becomes straightforward after using the series representation of the logarithm function, $\log (1-x) = \sum_{n=1}^{\infty} x^n/n$.

\section{Uncharged models}

We start our investigations by looking at uncharged models, i.e., we set $e \equiv 0$, which also means that the gauge field is completely decoupled, and we only need to focus on the fluctuations of $\Phi$. For the same reason, no wave function renormalization appears at the leading order, and in order to determine the flow of the effective action, we are free to evaluate (\ref{Eq:flow2}) in a constant background field of $\Phi$, which considerably simplifies the structure of (\ref{Eq:flow2}) in momentum space. The $(\Gamma^{(2)0}_{k,R})^{-1}$ matrix in Fourier space simply becomes (for $|q|<k$)
\bea
(\Gamma_{k,R}^{(2)0})^{-1}=(q^2+R_k)^{-1} \cdot{\bf 1} \equiv k^{-2}\cdot{\bf 1}.
\eea
After performing the $\tilde{\partial}_k$ differentiation, (\ref{Eq:flow2}) leads to (note that now $U_k\equiv V_k$)
\bea
\label{Eq:flow3}
k\partial_k \Gamma_k = \delta(0) \frac{k^{\cal D} \Omega_{\cal D}}{\cal D} \sum_{j=1}^{\infty} \left(\frac{-1}{k^2}\right)^j \Tr (V_k'')^j.
\eea
Here $\Omega_{\cal D}=\int_\Omega d\Omega/(2\pi)^{\cal D}$, and $\delta(0)$ is just a spacetime volume and is always canceled against a similar term in the lhs when evaluated in a constant background field. From the definition of $V_k$ [see (\ref{Eq:Gamma2}) and (\ref{Eq:basicG})], we have
\bea
\label{Eq:Vknomass}
V_k &=& g_{1,k} |\Tr(\Phi^\dagger \Phi)|^2 + g_{2,k} \Tr (\Phi^\dagger \Phi \Phi^\dagger \Phi),
\eea
where $Z_k=1$ is assumed, and we have set $\mu_k^2\equiv 0$. It has already been argued in \cite{fejos17}, and will be mentioned later that, $\mu_k^2\neq 0$ introduces gauge anomalies when calculating the flow of the wave function renormalization of the gauge field. We wish to avoid this problem in this study, and we also have in mind that signs of IR stable fixed points can be obtained also in massless schemes \cite{fejos16,zinnjustin}. We are, therefore, left with calculating the flow of $g_{1,k}$, $g_{2,k}$, and most importantly, as announced in the previous section, investigating under what circumstances the renormalization group flows close.

\subsection{Simple algebras}

First, we assume that the $\{T_a\}$ matrices span a simple Lie algebra; i.e., there are no two mutually commuting sets of generators [and thus obviously no $U(1)$ factors]. We are working in the fundamental representation, and thus, the product of two generators lies in the space of the algebra plus the identity,
\bea
\label{Eq:TT}
T_iT_j = \frac{N_T}{4}\delta_{ij}{\bf 1}+\frac12(d_{ijk}+if_{ijk})T_k,
\eea 
where ${\bf 1}$ is the unit matrix, $N_T=2/\Tr({\bf 1})$, and $d_{ijk}$ and $f_{ijk}$ are totally symmetric and antisymmetric structure constants, respectively. The reader finds the basics and useful identities of Lie algebras in Appendix A.
The traces in $V_k$ are evaluated as
\begin{subequations}
\label{Eq:inva}
\bea
\left(\Tr(\Phi^\dagger \Phi)\right)^2& =& \frac14(s^as^a+\pi^a\pi^a)^2, \\
\label{Eq:Trphi4}
\Tr(\Phi^\dagger \Phi\Phi^\dagger \Phi) &=&\frac{N_T}{2} |\Tr(\Phi^\dagger \Phi)|^2 \nonumber\\
&+&\frac{N_T}{2}(s^a s^a\pi^b\pi^b - (s^a\pi^a)^2) \nonumber\\
&+& \frac{1}{24}D_{abcd}(s^as^bs^cs^c+\pi^a\pi^b\pi^c\pi^d) \nonumber\\
&+&(\tilde{D}_{ab,cd}-D_{abcd}/4)s^as^b\pi^c\pi^d,
\eea
\end{subequations}
where we have introduced
\begin{subequations}
\bea
\label{Eq:D}
D_{abcd}&=&d_{abm}d_{cdm}+d_{adm}d_{bcm}+d_{acm}d_{bdm}, \\
\label{Eq:Dtil}
\tilde{D}_{ab,cd}&=&d_{abm}d_{cdm}.
\eea
\end{subequations}
It is worth to list the multiplication rule between these tensors. Using the notation $(D*D)_{abcd}=D_{abij}D_{ijcd}$, we get (see Appendix A for useful formulas)
\begin{subequations}
\label{Eq:Dtable}
\bea
(D*D)_{abcd}&=&\big(6N_T(d-2)-10C_2(A)\big)\tilde{D}_{ab,cd} \nonumber\\
&+&\big((d-3)N_T-2C_2(A)\big)D_{abcd} \nonumber\\
&+&N_T((d-1)N_T-C_2(A))\nonumber\\
&\times&(2\delta_{ab}\delta_{cd}+\delta_{ad}\delta_{bc}+\delta_{ac}\delta_{bd}), \\
(D*\tilde{D})_{abcd}&\equiv& (\tilde{D}*D)_{abcd} \nonumber\\
&=&\big(N_T(3d-5)-4C_2(A)\big)\tilde{D}_{ab,cd}, \\
(\tilde{D}*\tilde{D})_{abcd}&=&\big(N_T(d-1)-C_2(A)\big)\tilde{D}_{abcd}.
\eea
\end{subequations}
Here, the notation $C_2(A)$ represents the value of the $T^2$ Casimir operator in the adjoint representation, i.e., $(T_kT_k|_A)_{ij}=C_2(A)\delta_{ij}$, or alternatively, $f_{ilk}f_{jlk}=C_2(A)\delta_{ij}$.

In order to check whether $\Gamma_k$ respects the form of the classical action, we have to evaluate the $j=2$ term in the expansion. Higher order terms produce operators that are not relevant from a renormalization point of view, since they are absent in the classical action as their coefficients have to go to zero in the continuum limit. The $j=1$ term, in turn, is not interesting as it can be easily shown to only produce contributions that are proportional to $\Tr(\Phi^\dagger \Phi)$. Therefore, we only need to evaluate $\Tr (V_k'' V_k'')$, where $V_k''$ can be thought of as a $2\times 2$ matrix, with $d\times d$ matrices in each entry (here, $d$ denotes the number of generators), in accordance with differentiations with respect to the fields $s^a$ or $\pi^a$,
\bea
\label{Eq:Vkdpr}
V_k''&=&
\begin{pmatrix}
V_{k,ss}'' & V_{k,s\pi}'' \\
V_{k,\pi s}'' & V_{k,\pi \pi}'' \\
\end{pmatrix}
\nonumber\\
&\equiv&
\begin{pmatrix}
g_{1,k}A_{ss}+g_{2,k}B_{ss} &g_{1,k}A_{s\pi}+g_{2,k}B_{s\pi} \\
g_{1,k}A_{\pi s}+g_{2,k}B_{\pi s} & g_{1,k}A_{\pi\pi}+g_{2,k}B_{\pi\pi} \\
\end{pmatrix},
\eea
where we have introduced the following matrices: 
\bea
\label{Eq:AB}
A=[|\Tr(\Phi^\dagger \Phi)|^{2}]'',\quad B=[\Tr(\Phi^\dagger \Phi \Phi^\dagger \Phi)]''. 
\eea
For the sake of helping understand the notations, e.g. $(B_{s\pi})_{ab}=\partial^2\Tr(\Phi^\dagger \Phi \Phi^\dagger \Phi)/\partial s^a \partial \pi^b$. Then, we get
\bea
\label{Eq:TrVVform}
\Tr&&\!\!\!\!\!\!\!(V_k'' V_k'')=\nonumber\\
&&g_{1,k}^2[\Tr A_{ss}^2+\Tr A_{\pi\pi}^2+2\Tr(A_{s\pi}A_{\pi s})] \nonumber\\
&+&2g_{1,k}g_{2,k}[\Tr (A_{ss}B_{ss})+\Tr (A_{\pi\pi}B_{\pi\pi})+\nonumber\\
&&\hspace{1.4cm}+2\Tr(A_{s\pi}B_{\pi s})] \nonumber\\
&+&g_{2,k}^2[\Tr B_{ss}^2+\Tr B_{\pi\pi}^2+2\Tr(B_{s\pi}B_{\pi s})].
\eea
Using the multiplication table (\ref{Eq:Dtable}), evaluation of the traces is straightforward, but tedious, as one needs to assume the most general background of $s^a$ and $\pi^a$. The reader is referred to the Appendixes for details. We get
\bea
\label{Eq:TrVV}
\Tr (V_k'' V_k'') &=& 4(2d+8) g_{1,k}^2  |\Tr (\Phi^\dagger \Phi)|^2 \nonumber\\
+&&\!\!\!\!\!\!\! 16g_{1,k}g_{2,k} \left[dN_T |\Tr (\Phi^\dagger \Phi)|^2+3 \Tr(\Phi^\dagger \Phi \Phi^\dagger \Phi) \right] \nonumber\\
+&&\!\!\!\!\!\!\! g_{2,k}^2 \Big[4N_T(3C_2(A)+8N_T) |\Tr (\Phi^\dagger \Phi)|^2\nonumber\\
+&&\!\!\!\!\!\!(20N_T(d-1)-24C_2(A))\Tr(\Phi^\dagger \Phi \Phi^\dagger \Phi) \nonumber\\
+&&\!\!\!\!\!\!12N_Ts^as^b\pi^c\pi^d(3\tilde{D}_{ab,cd}-D_{abcd}) \nonumber\\
-&&\!\!\!\!\!\!22N_T^2|\Tr(\Phi\Phi)|^2\Big],
\eea
which shows that the RG flow does not respect the form of the classical potential, as not only terms built up by $\Tr (\Phi^\dagger \Phi)$ or $\Tr(\Phi^\dagger \Phi \Phi^\dagger\Phi)$ are formed. This is one of the important results of the paper, showing that for a general simple Lie algebra, the field theory defined in (\ref{Eq:basicL}) containing two quartic couplings is not one-loop closed.

\subsection{Simple algebras: $SU(2)$}

There are some exceptions, though. Take, for example $SU(2)$. Then the last line of (\ref{Eq:TrVV}) is identically zero [for $SU(2)$ $d_{abc}\equiv 0$], and one uses the identity,
\bea
\Tr(\Phi^\dagger \Phi \Phi^\dagger\Phi)\big|_{SU(2)}&=&|\Tr (\Phi^\dagger \Phi)|^2\big|_{SU(2)}\nonumber\\
&-&|\Tr(\Phi\Phi)|^2/2|_{SU(2)}
\eea
to get
\bea
\label{Eq:TrVVsu2}
\Tr (V_k'' V_k'')|_{SU(2)} &=& 56 g_{1,k}^2  |\Tr (\Phi^\dagger \Phi)|^2 \nonumber\\
&&\!\!\!\!\!\!\!\!\!\!\!\!\!\!\!\!\!\!\!\!\!\!\!\!\! +48g_{1,k}g_{2,k} \left[|\Tr (\Phi^\dagger \Phi)|^2+ \Tr(\Phi^\dagger \Phi \Phi^\dagger \Phi) \right] \nonumber\\
&&\!\!\!\!\!\!\!\!\!\!\!\!\!\!\!\!\!\!\!\!\!\!\!\!\! +12g_{2,k}^2 \big[|\Tr (\Phi^\dagger \Phi)|^2+3\Tr(\Phi^\dagger \Phi \Phi^\dagger \Phi)\big],
\eea
where we also used that $N_T=1$, $C_2(A)=2$, $d=3$. Equation (\ref{Eq:TrVVsu2}) shows that the LPA of a $SU(2)$ theory is one-loop closed, and the flows of the couplings can be read off by combining (\ref{Eq:TrVVsu2}) with (\ref{Eq:flow3}),
\begin{subequations}
\bea
k\partial_k g_{1,k}|_{SU(2)}&=&\frac{k^{{\cal D}-4}\Omega_{\cal D}}{{\cal D}}(56g_{1,k}^2+48g_{1,k}g_{2,k}+12g_{2,k}^2), \nonumber\\
\\
k\partial_k g_{2,k}|_{SU(2)}&=&\frac{k^{{\cal D}-4}\Omega_{\cal D}}{{\cal D}}(48g_{1,k}g_{2,k}+36g_{2,k}^2).
\eea
\end{subequations}
\subsection{Simple algebras: $SU(3)$}

We can now try $SU(3)$. First, we make use of
\bea
D_{abcd}|_{SU(3)}=\frac13(\delta_{ab}\delta_{cd}+\delta_{ac}\delta_{bd}+\delta_{ad}\delta_{bc}),
\eea
\newline
and then from (\ref{Eq:Trphi4}) express $\tilde{D}_{ab,cd}s^as^b\pi^c\pi^d$ as
\bea
&&\tilde{D}_{ab,cd}s^as^b\pi^c\pi^d|_{SU(3)}=\Tr(\Phi^\dagger \Phi \Phi^\dagger \Phi)|_{SU(3)}\nonumber\\
&&-\frac12|\Tr (\Phi^\dagger \Phi)|^2|_{SU(3)}-\frac16s^as^a\pi^b\pi^b+\frac12(s^a\pi^a)^2
\eea
using that $N_T=2/3$, $C_2(A)=3$, $d=8$.
This helps a bit, but not quite, as
\bea
\label{Eq:TrVVsu3}
\Tr &&\!\!\!\!\!\!\!(V_k'' V_k'')|_{SU(3)}=96g^2_{1,k}|\Tr (\Phi^\dagger \Phi)|^2\nonumber\\
&+&g_{1,k}g_{2,k}\Big[\frac{256}{3}|\Tr (\Phi^\dagger \Phi)|^2+48\Tr (\Phi^\dagger \Phi\Phi^\dagger \Phi)\Big] \nonumber\\
&+&\frac{g^2_{2,k}}{9}\Big[176|\Tr (\Phi^\dagger \Phi)|^2+408\Tr(\Phi^\dagger \Phi \Phi^\dagger \Phi)\nonumber\\
&&\hspace{0.8cm}-28|\Tr(\Phi\Phi)|^2\Big],
\eea
which shows that the flow, again, does not close, as $|\Tr(\Phi\Phi)|^2$ is absent in (\ref{Eq:Vknomass}). But then, one can try to build up another theory based on the $SU(3)$ structure, which does include the new term in the $V_k$ potential,
\bea
\label{Eq:Vksu3}
V_k &=& g_{1,k} |\Tr(\Phi^\dagger \Phi)|^2 + g_{2,k} \Tr (\Phi^\dagger \Phi \Phi^\dagger \Phi) \nonumber\\
&-& g_{3,k} \big(|\Tr(\Phi\Phi)|^2-|\Tr(\Phi^\dagger \Phi)|^2\big),
\eea
where, only out of computational convenience, we have separated $|\Tr(\Phi^\dagger \Phi)|^2$ from the new operator. The $V_k''$ matrix changes as
\begin{widetext}
\bea
V_k''&=&
\begin{pmatrix}
V_{k,ss}'' & V_{k,s\pi}'' \\
V_{k,\pi s}'' & V_{k,\pi \pi}'' \\
\end{pmatrix}=
\begin{pmatrix}
g_{1,k}A_{ss}+g_{2,k}B_{ss}+g_{3,k}C_{ss} &g_{1,k}A_{s\pi}+g_{2,k}B_{s\pi} +g_{3,k}C_{s\pi}\\
g_{1,k}A_{\pi s}+g_{2,k}B_{\pi s}+g_{3,k}C_{\pi s}& g_{1,k}A_{\pi\pi}+g_{2,k}B_{\pi\pi}+g_{3,k}C_{\pi \pi} \\
\end{pmatrix},
\eea
where the $A$ and $B$ matrices are given, again, by (\ref{Eq:AB}), and
\bea
\label{Eq:defC}
C =\big(|\Tr(\Phi^\dagger \Phi)|^2-|\Tr(\Phi\Phi)|^2\big)'' \equiv(s^as^a\pi^b\pi^b-(s^a\pi^a)^2)''
\eea
where the double primes refer, again, to field differentiations; see the terminology below (\ref{Eq:AB}). $\Tr(V_k''V_k'')|_{SU(3)}$ gets the following correction:
\bea
\Delta\!\Tr(V_k''V_k'')&=&g_{3,k}^2[\Tr C_{ss}^2+\Tr C_{\pi\pi}^2+2\Tr(C_{s\pi}C_{\pi s})] +2g_{1,k}g_{3,k}[\Tr (A_{ss}C_{ss})+\Tr (A_{\pi\pi}C_{\pi\pi})+2\Tr(A_{s\pi}C_{\pi s})] \nonumber\\
&+&2g_{2,k}g_{3,k}[\Tr (B_{ss}C_{ss})+\Tr (B_{\pi\pi}C_{\pi\pi})+2\Tr(B_{s\pi}C_{\pi s})],
\eea
and after some algebra we get (see also Appendix B)
\bea
\label{Eq:DTrVV}
\Delta\! \Tr(V_k''V_k'')&=&g_{3,k}^2\Big[96|\Tr(\Phi^\dagger \Phi)|^2+16|\Tr(\Phi\Phi)|^2\Big]+g_{1,k}g_{3,k}\Big[160|\Tr(\Phi^\dagger \Phi)|^2-48|\Tr(\Phi\Phi)|^2)\Big] \nonumber\\
&+&g_{2,k}g_{3,k}\Big[\frac{160}{3}|\Tr(\Phi^\dagger \Phi)|^2+80\Tr (\Phi^\dagger \Phi \Phi^\dagger \Phi)-\frac{32}{3}|\Tr(\Phi\Phi)|^2\Big],
\eea
\end{widetext}
which shows that this theory, defined via (\ref{Eq:Vksu3}), is one-loop closed, and the coupling flows are [see (\ref{Eq:DTrVV}) and (\ref{Eq:TrVVsu3})]
\begin{subequations}
\bea
k\partial_k g_{1,k}&=&\frac{k^{{\cal D}-4}\Omega_{\cal D}}{{\cal D}}\Big[96g_{1,k}^2+\frac{256}{3}g_{1,k}g_{2,k}+\frac{148}{9}g_{2,k}^2\nonumber\\
&+&112g_{3,k}^2+112g_{1,k}g_{3,k}+\frac{128}{3}g_{2,k}g_{3,k}\Big], \\
k\partial_k g_{2,k}&=&\frac{k^{{\cal D}-4}\Omega_{\cal D}}{{\cal D}}\Big[48g_{1,k}g_{2,k}+\frac{136}{3}g_{2,k}^2\nonumber\\
&&\hspace{1.5cm}+80g_{2,k}g_{3,k}\Big], \\
k\partial_k g_{3,k}&=&\frac{k^{{\cal D}-4}\Omega_{\cal D}}{{\cal D}}\Big[\frac{28}{9}g_{2,k}^2-16g_{3,k}^2+48g_{1,k}g_{3,k}\nonumber\\
&&\hspace{1.5cm}+\frac{32}{3}g_{2,k}g_{3,k}\Big].
\eea
\end{subequations}
\subsection{Simple algebras extended with a $U(1)$ factor}

One expects that by extending the potential with more operators, it might be possible to build up one-loop closed theories (from a RG point of view) in $SU(n)$-like theories. We are still interested, however, if the original construction (\ref{Eq:Vknomass}) can lead to closed flows. Here, we show that it is sufficient to extend any simple Lie algebra with one $U(1)$ factor for that. It turns out that the closed RG flow boils down to that if one $U(1)$ factor is included, not only the commutator, but also the anticommutator belongs to the algebra itself [this is not the case for simple algebras, see (\ref{Eq:TT})]. Since for any matrix $\Phi_1$ and $\Phi_2$,
\bea
\Phi_1 \cdot \Phi_2 = \frac12[\Phi_1,\Phi_2]+\frac12\{\Phi_1,\Phi_2\},
\eea
the algebra is closing not only with respect to the Lie bracket but also to matrix multiplication. Denoting the new generator, which generates the additional $U(1)$ factor, by $T_0\equiv \sqrt{N_T}/2\cdot {\bf 1}$, one generalizes (\ref{Eq:TT}) to
\bea
T_iT_j = \frac12 (d_{ijk}+if_{ijk})T_k, \\ \nonumber
\eea
where $d_{ij0}=\sqrt{N_T}\delta_{ij}$, and $f_{ij0}\equiv 0$. The procedure is the same as before, first we calculate the following traces:
\begin{subequations}
\bea
|\Tr(\Phi^\dagger \Phi)|^2& =& \frac14(s^as^a+\pi^a\pi^a)^2, \\
\label{Eq:Trphi4b}
\Tr(\Phi^\dagger \Phi\Phi^\dagger \Phi) &=&\frac{1}{24}D_{abcd}(s^as^bs^cs^c+\pi^a\pi^b\pi^c\pi^d) \nonumber\\
&+&(\tilde{D}_{ab,cd}-D_{abcd}/4)s^as^b\pi^c\pi^d,
\eea
\end{subequations}
where the second expression (\ref{Eq:Trphi4b}) looks significantly simpler than that of the case of a simple algebra (\ref{Eq:Trphi4}). The $D$ and $\tilde{D}$ tensors are defined exactly as in (\ref{Eq:D}) and (\ref{Eq:Dtil}), but note that now summations go through all indices, including $m=0$. The multiplication table becomes

\begin{widetext}
\begin{subequations}
\bea
\label{Eq:Dtable2}
(D*D)_{abcd}&=&\big(6N_Td-8C_2(A)\big)\tilde{D}_{ab,cd} +(N_Td-2C_2(A))D_{abcd}+C_2(A)N_T(4\delta_{ab}\delta_{cd}+\delta_{ac}\delta_{bd}+\delta_{ad}\delta_{bc}) \nonumber\\
&+&2C_2(A)\sqrt{N_T}(\delta_{a0}d_{bcd}+\delta_{b0}d_{acd}+\delta_{c0}d_{abd}+\delta_{d0}d_{abc}),\\
(D*\tilde{D})_{abcd}&=&\big(3N_Td-4C_2(A)\big)\tilde{D}_{ab,cd}
+2C_2(A)N_T\delta_{ab}\delta_{cd} +C_2(A)\sqrt{N_T}[\delta_{a0}d_{bcd}+\delta_{b0}d_{acd}], \\
(\tilde{D}*D)_{abcd}&=&\big(3N_Td-4C_2(A)\big)\tilde{D}_{ab,cd}
+2C_2(A)N_T\delta_{ab}\delta_{cd}+C_2(A)\sqrt{N_T}[\delta_{c0}d_{abd}+\delta_{d0}d_{abc}], \\
(\tilde{D}*\tilde{D})_{abcd}&=&\big(N_Td-C_2(A)\big)\tilde{D}_{abcd}+C_2(A)N_T\delta_{ab}\delta_{cd}.
\eea
\end{subequations}
\end{widetext}
A long and tedious calculation of the traces of the terms including the $A$ and $B$ matrices [see definitions, again, in (\ref{Eq:AB})] leads to
\bea
\Tr&&\!\!\!\!\!\!\!\!(V_k''V_k'')=g_{1,k}^28(d+4)|\Tr(\Phi^\dagger \Phi)|^2\nonumber\\
&+&g_{1,k}g_{2,k}\Big[16N_Td|\Tr(\Phi^\dagger \Phi)|^2+48\Tr(\Phi^\dagger \Phi\Phi^\dagger\Phi)\Big]\nonumber\\
&+&g_{2,k}^2\Big[12N_TC_2(A)|\Tr(\Phi^\dagger \Phi)|^2\nonumber\\
&&\hspace{0.7cm}+(20N_Td-24C_2(A))\Tr(\Phi^\dagger \Phi\Phi^\dagger\Phi)\Big].
\eea
This is another important result showing that by including into the algebra one $U(1)$ factor, the renormalization group flows at one-loop always close and thus these theories are consistent. The scale dependence of the couplings are described by
\bea
\label{Eq:finalg1k}
k\partial_k g_{1,k}&=&\frac{k^{{\cal D}-4}\Omega_{\cal D}}{{\cal D}}\Big[8(d+4)g_{1,k}^2+16N_Tdg_{1,k}g_{2,k}\nonumber\\
&&\hspace{1.6cm}+12C_2(A)N_Tg_{2,k}^2\Big], \\
\label{Eq:finalg2k}
k\partial_k g_{2,k}&=&\frac{k^{{\cal D}-4}\Omega_{\cal D}}{{\cal D}}\Big[48g_{1,k}g_{2,k}\nonumber\\
&&\hspace{1.6cm}+(20N_Td-24C_2(A))g_{2,k}^2\Big].
\eea
For example, in the case of $SU(n)\otimes U(1)\simeq U(n)$, $N_t=2/n$, $d=n^2$, $C_2(A)=n$, and we get
\bea
\label{Eq:finalg1kun}
k\partial_k g_{1,k}&=&\frac{k^{{\cal D}-4}\Omega_{\cal D}}{{\cal D}}\Big[8(n^2+4)g_{1,k}^2+32ng_{1,k}g_{2,k}+24g_{2,k}^2\Big]\nonumber\\
\\
\label{Eq:finalg2kun}
k\partial_k g_{2,k}&=&\frac{k^{{\cal D}-4}\Omega_{\cal D}}{{\cal D}}\Big[48g_{1,k}g_{2,k}+16ng_{2,k}^2\Big],
\eea
which agree with the well-known result of Pisarski and Wilczek \cite{pisarski84}.

\section{Charged models}

Now that we identified what scalar theories are one-loop closed in a RG flow sense, we take into account the gauge field and the charge, $e \neq 0$. We assume the $\Phi$ field lies in a simple Lie algebra with an additional $U(1)$ factor. Our goal in this section is to obtain the corrections to the coupling flows coming from the charge, and the flow of the charge itself.

Once the charge is taken into account, the wave function renormalization $Z_k$ of the $\Phi$ field cannot be dropped, as the charge produces a nonzero contribution even at leading order. Therefore, formulas (\ref{Eq:finalg1k}) and (\ref{Eq:finalg2k}) are still valid, but we need to make the substitutions $g_{1,k} \rightarrow Z_k^2g_{1,k}$, $g_{2,k} \rightarrow Z_k^2 g_{2,k}$, and take into account an additional $1/Z_k^2$ factor in the rhs of (\ref{Eq:finalg1k}) and (\ref{Eq:finalg2k}) coming from the propagator,
\bea
\label{Eq:g1ka}
k\partial_k (Z_k^2g_{1,k})&=&Z_k^2\frac{k^{{\cal D}-4}\Omega_{\cal D}}{{\cal D}}\Big[8(d+4)g_{1,k}^2+16N_Tdg_{1,k}g_{2,k}\nonumber\\
&&\hspace{1.6cm}+12C_2(A)N_Tg_{2,k}^2\Big] \nonumber\\
&&\hspace{1.6cm}+{\cal O}(e_k^2g_{1,k},e_k^2g_{2,k},e_k^4),\\
\label{Eq:g2ka}
k\partial_k (Z_k^2g_{2,k})&=&Z_k^2\frac{k^{{\cal D}-4}\Omega_{\cal D}}{{\cal D}}\Big[48g_{1,k}g_{2,k}\nonumber\\
&&\hspace{1.6cm}+(20N_Td-24C_2(A))g_{2,k}^2\Big]\nonumber\\
&&\hspace{1.6cm}+{\cal O}(e_k^2g_{1,k},e_k^2g_{2,k},e_k^4).
\eea
Before we proceed, it is worth to reformulate (\ref{Eq:basicG}) as [note, again, that, $\Phi=(s^a+i\pi^a)T_a$]
\bea
\label{Eq:Lkref}
L_k &=&  Z_{A,k}\frac{{\cal A}_i}{2}\left(-\partial^2 \delta_{ij} + (1-\xi_k^{-1})\partial_i\partial_j\right){\cal A}_j \nonumber\\
&+&\frac{Z_k}{2}(\partial_i s^a\partial_is^a +\partial_i \pi^a\partial_i\pi^a) \nonumber\\
&+&Z_{e,k}e{\cal A}_i(s^a\partial_i\pi^a-\pi^a\partial_i s^a)\nonumber\\
&+&\frac{Z_{e,k}^2}{Z_k}\frac{e^2}{2}{\cal A}_i{\cal A}_i(s^as^a+\pi^a\pi^a)+V_k[s^a,\pi^a].
\eea
This form is more useful for reading off Feynman rules.

\subsection{Scalar wave function renormalization}
We start analyzing the charged flows by calculating the flow of the scalar wave function renormalization $Z_k$. Note that the only coupling that contributes to the flow of $Z_k$ is the third term in (\ref{Eq:Lkref}), which does not depend at all on the Lie algebra structure, only the number of $s^a$ and $\pi^a$ fields matters. Therefore, a parallel calculation can be done as in \cite{fejos17}, which is essentially the same theory as (\ref{Eq:Lkref}), but in the former the potential did not contain the second quartic coupling $g_{2,k}$.

Comments on the regularization scheme is now in order. In \cite{fejos17} it was shown that different regulators lead to different predictions for the $Z_k$ wave function renormalization factor. The gauge fixing, used earlier in \cite{fejos17}, the $R_\xi$ choice, allowed us to compute $k\partial_k Z_k$ fairly easy, since the propagator matrix was diagonal in momentum space even when the fields were not homogeneous. (This gauge choice also had a disadvantage, which we will come back to in Sec. IVD.) The eigenvalues of the inverse propagator matrix were of the form $\sim[q^2 + (...)/q^2]$, and one had the choice of replacing via the regulator all $q$ dependence with $k$ (R2 regulator) or only the Gaussian part (i.e., $\sim q^2$, R1 regulator). The latter led to better convergence properties, and one concluded that this is a more legitimate choice. However, since we are now working in the ordinary covariant gauge, and thus, the inverse propagator is nondiagonal in momentum space, it is highly nontrivial (at least at first sight) what regulator corresponds to the preferred choice (i.e., R1 in the $R_\xi$ gauge). We show here that the regulator we are looking for in this gauge is remarkably simple: apart from a small catch one need not regulate the gauge field, ${\cal A}_i$, only the scalars, i.e., $s^a$ and $\pi^a$. We will not discuss it in detail, but it turns out that the other choice, corresponding to R2, is also simple; there one associates regulators, as usual, to all dynamical variables, ${\cal A}_i$, $s^a$, and $\pi^a$.

\begin{figure}
\includegraphics[bb = 350 500 380 550,scale=0.9,angle=0]{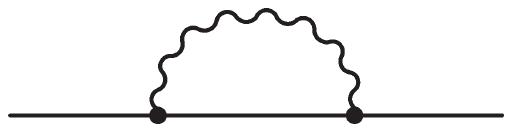}
\caption{The diagram that is responsible for the flow of the scalar wave function renormalization factor $Z_k$. Solid lines refer to the scalar (and pseudoscalar) fields, while the wiggly one is the gauge propagator. No tensor structure is indicated explicitly. In the regularization proposed here, the gauge propagator is not regulated.}
\end{figure}  

There are at least two ways to perform the calculation of $k\partial_kZ_k$, leading to identical results. One way is to brute force calculate the corresponding terms in (\ref{Eq:flow2}), but it is simpler to use diagrammatics. Since $\Tr \log$ of the propagator is the sum of one-loop diagrams, one needs to evaluate only one graph, shown in Fig. 1. Not regulating the gauge field, we get the following contribution into $k\partial_k \Gamma_k$:
\bea
\label{Eq:Zflow}
k\partial_k \Gamma_k \hspace{0.15cm}&\Leftarrow&\hspace{0.15cm} -e^2k\tilde{\partial}_k\frac{Z_{e,k}^2}{2Z_{A,k}Z_k}\int_{p} (s^a_{-p}s^a_p+\pi^a_{-p}\pi^a_p)\nonumber\\
&&\hspace{0.2cm}\times\int_q\Big[\delta^{\alpha \beta}-(1-\xi_k)\frac{(p-q)^\alpha(p-q)^\beta}{(p-q)^2}\Big]\nonumber\\
&&\hspace{0.2cm}\times\frac{(p+q)^\alpha(p+q)^\beta}{{\big(q^2+R_k(q)\big)(q-p)^2}}. 
\eea
Neglecting anomalous dimensions and considering only the ${\cal O}(p^2)$ terms in the $q$ integral (the mass flow, i.e., terms with $\sim p^0$, is not interesting), we arrive at
\bea
p^2k\partial_k Z_k&=&e^2\frac{4Z_{e,k}^2}{Z_{A,k}Z_k}\int_q \frac{k\partial_k R_k(q)}{(q^2+R_k(q))^2}\nonumber\\
&\times&\frac{1-\xi_k-x^2+3\xi_kx^2}{q^2}p^2,
\eea
where $x=\hat{p}\cdot \hat{q}$. As announced in Sec. II, we work with the optimal regulator, $R_k(q)=(k^2-q^2)\Theta(k^2-q^2)$, and get
\bea
\label{Eq:dkZk}
k\partial_k Z_k = \frac{k^{{\cal D}-4}\Omega_{\cal D}}{{\cal D}}\frac{8e_k^2Z_k}{({\cal D}-2)}({\cal D}-1+\xi_k(3-{\cal D})),
\eea
where we have also introduced the flowing charge; see above (\ref{Eq:flow1}), $e_k=eZ_{e,k}/Z_kZ^{1/2}_{A,k}$.

Note that (\ref{Eq:dkZk}) does not agree with the corresponding result of \cite{fejos17}, but this is only because we are working in a different gauge. We will see that once combined with the flow $k\partial_k(g_{1,k}Z_k^2)$ [see (\ref{Eq:finalg1k})], the coupling flow $k\partial_k g_{1,k}$ is consistent with that of \cite{fejos17} (note that there no second coupling, $g_{2,k}$, was present).

\begin{figure}
\includegraphics[bb = 210 480 380 550,scale=0.8,angle=0]{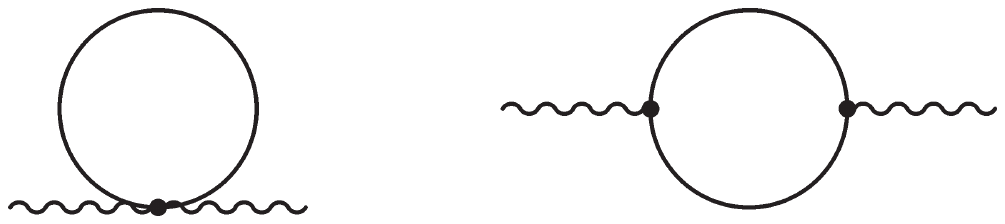}
\caption{Diagrams that contribute to the flow of the gauge wave function renormalization factor, $Z_{A,k}$. Solid lines mean scalar (and pseudoscalar) propagators, while the wiggly ones refer to the gauge field.}
\end{figure}  

\subsection{Gauge wave function renormalization}

The calculation of the gauge wave function renormalization is completely identical to that of \cite{fejos17}. Two diagrams need to be taken into account; see Fig. 2. We just review the result here: the flow of $Z_{A,k}$ is
\bea
\label{Eq:dkZAk}
k\partial_k Z_{A,k}=-\frac{k^{{\cal D}-4}\Omega_{\cal D}}{{\cal D}}\frac{16de_k^2}{{\cal D}+2}Z_{A,k},
\eea
and it turns out that even though one expects from the ansatz (\ref{Eq:basicG}) the combination
\bea
\label{Eq:diagexp}
k\partial_k \Big[Z_{A,k} \int \frac{{\cal A}_i}{2} (-\partial^2\delta_{ij}+(1-\xi_k^{-1})\partial_i\partial_j) {\cal A}_j \Big]
\eea
to emerge in the rhs of (\ref{Eq:flow1}), one actually gets from the two diagrams above
\bea
\label{Eq:diagsum}
k\partial_k \Big[Z_{A,k}\int \frac{{\cal A}_i}{2} (-\partial^2\delta_{ij}+\frac{{\cal D}-2}{2}\partial_i\partial_j) {\cal A}_j \Big].
\eea
If ${\cal D}\neq 4$, this selects only one gauge parameter that is consistent,
\bea
\label{Eq:xicons}
\xi_k \equiv 2/(4-{\cal D}).
\eea
For ${\cal D}=4$, any choice is allowed, as long as the gauge fixing parameter follows the flow of the gauge wave function renormalization, i.e., $\xi_k \sim Z_{A,k}$, in accordance with perturbation theory. We will see that in this particular dimension the $\beta$-functions do not depend on the actual value of $\xi_k$ after all. We also note that the induced mass of the gauge field, which comes from the momentum independent part of the two diagrams of Fig. 2, is completely dropped. It has been shown that neglecting it is not of any concern, since once adjusted at the UV scale, this term completely dies out in the IR and has no relevance \cite{gies12,fejos17}.

Also note that, as announced in the beginning of Sec. III, the scalar mass was set to zero, $\mu_k\equiv 0$. Had we not imposed this requirement, the result (\ref{Eq:diagsum}) would not be compatible with (\ref{Eq:diagexp}) for any gauge fixing parameter. It would be interesting to further analyze the source of this violation of gauge symmetry, but here we leave it for future studies.

\subsection{Charge corrections to the coupling flows}

We make use of diagrammatics once again; see Fig. 3. The first diagram is already done, as the whole previous section was devoted to analyze that very piece; the results are summarized in (\ref{Eq:finalg1k}) and (\ref{Eq:finalg2k}). (Note that we do not differentiate the tensor structure in the diagrams, we only draw them for topological distinction.) The second piece is responsible for the ${\cal O}(g_{1,k}e_k^2)$ and ${\cal O}(g_{2,k}e_k^2)$ terms in the coupling flows. These type of terms lead to the following contribution into $k\partial_k \Gamma_k$:
\bea
\label{Eq:ge2}
k\partial_k \Gamma_k&\Leftarrow&k\tilde{\partial}_k \frac{e^2Z_{e,k}^2}{6Z_kZ_{A,k}}
\int_x (s^as^b(V_{k,ss}'')_{ab}+\pi^a\pi^b(V_{k,\pi\pi}'')_{ab}\nonumber\\
&&\hspace{2.2cm}+\pi^as^b(V''_{k,\pi s})_{ab}+s^a\pi^b(V''_{k,s \pi})_{ab}) \nonumber\\
&\times&\int_q\frac{1}{(q^2+R_k(q))^2}\frac{1}{q^2}\Big(\delta^{\alpha \beta}-(1-\xi_k)\frac{q^\alpha q^\beta}{q^2}\Big)q^{\alpha}q^{\beta}, \nonumber\\
\eea
where the $V_k''$ matrix can be built up from the $A$ and $B$ matrices, see definitions in (\ref{Eq:Vkdpr}) and (\ref{Eq:AB}), and useful formulas in Appendix B. Neglecting the anomalous dimension in the rhs of (\ref{Eq:ge2}), we get
\bea
\label{Eq:ge2final}
\int_x &&\!\!\!\!\!\Big(-\frac{\Omega_{\cal D}}{{\cal D}}k^{{\cal D}-4}8\xi_k e_k^2Z_k^2\Big)\nonumber\\
&\times&\Big(g_{1,k}|\Tr(\Phi^\dagger \Phi)|^2+g_{2,k}\Tr(\Phi^\dagger \Phi\Phi^\dagger \Phi)\Big). 
\eea
This result shows that only those four point operators appeared that are allowed by the Lagrangian, therefore, the RG flow is closed.

\begin{figure}
\includegraphics[bb = 140 370 380 550,scale=0.8,angle=0]{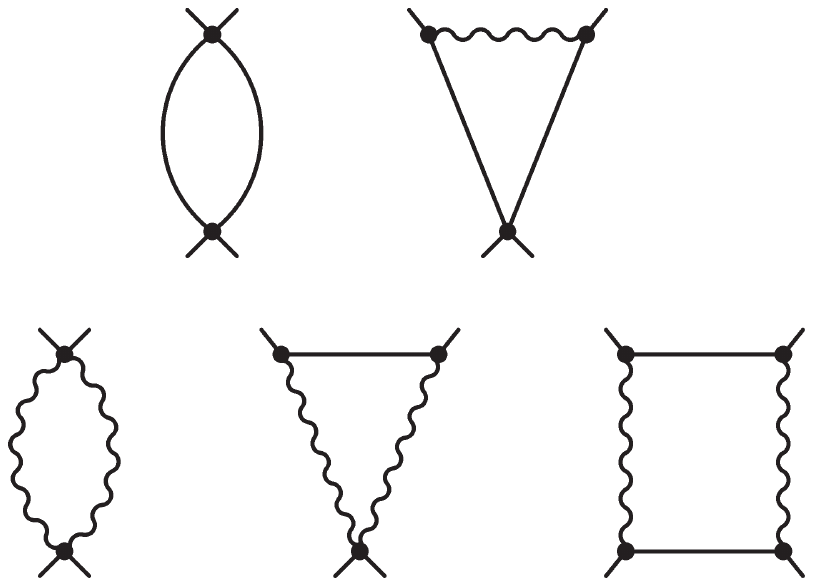}
\caption{Diagrams (with zero external momenta) that contribute to the flows of the coupling constants $g_{1,k}$ and $g_{2.k}$. Solid lines refer to the scalar (and pseudoscalar) fields, while the wiggly one is the gauge propagator. No tensor structure is indicated explicitly.}
\end{figure}  

Finally, we analyze the ${\cal O}(e_k^4)$ contribution to the coupling flows. The last three diagrams of Fig. 3 need to be evaluated. One immediately notes that if we wish to stick to the regularization, where the gauge propagators are not regulated, then we run into a divergence coming from the first diagram of the second line in Fig. 3. We have no choice but to restore the gauge regulator, and then the diagrams lead to the following contributions to $k\partial_k \Gamma_k$, respectively:
\begin{itemize}
\item $k\tilde{\partial}_k\Big(-\frac{Z_{e,k}^4e^4}{4Z_k^2Z_{A,k}^2}\Big)\int_x (s^as^a+\pi^a\pi^a)^2 \int_q\frac{1}{(q^2+R_k(q))^2}$ \newline
$\times \Big[\delta^{\alpha \beta}-(1-\xi_k)\frac{q^\alpha q^\beta}{q^2}\Big]\Big[\delta^{\beta \alpha}-(1-\xi_k)\frac{q^\beta q^\alpha}{q^2}\Big],$
\item $k\tilde{\partial}_k \frac{Z_{e,k}^4e^4}{2Z_k^2Z_{A,k}^2}\int_x (s^as^a+\pi^a\pi^a)^2 \int_q\frac{1}{(q^2+R_k(q))^3}$ \newline
$\times \Big[\delta^{\alpha \beta}-(1-\xi_k)\frac{q^\alpha q^\beta}{q^2}\Big]\Big[\delta^{\beta \gamma}-(1-\xi_k)\frac{q^\beta q^\gamma}{q^2}\Big]q^{\gamma}q^{\alpha},$
\item $k\tilde{\partial}_k \Big(-\frac{Z_{e,k}^4e^4}{4Z_k^2Z_{A,k}^2}\Big)\int_x (s^as^a+\pi^a\pi^a)^2 \int_q\frac{1}{(q^2+R_k(q))^4}$ \newline
$\times \Big\{\Big[\delta^{\alpha \beta}-(1-\xi_k)\frac{q^\alpha q^\beta}{q^2}\Big]q^{\alpha}q^{\beta}\Big\}^2$.
\end{itemize}
The sum of these three diagrams turn out to be
\bea
\label{Eq:e4inter}
\int_x &&\Big(\frac{4}{{\cal D}}({\cal D}-1)+\xi_k^2\Big[\frac{4}{{\cal D}}-\frac{12}{{\cal D}+2}+\frac{8}{{\cal D}+4}\Big]\Big)\nonumber\\
&&\hspace{0.1cm}\times \Omega_{\cal D} k^{{\cal D}-4} Z_k^2 e_k^4|\Tr(\Phi^\dagger \Phi)|^2.
\eea
The coefficient of the $\xi_k^2$ term does not cancel, except for ${\cal D}=4$. Had it canceled, it would have led to a compatible result with \cite{fejos17} in the $R_\xi$ gauge. The problem here is that the momentum dependent scalar (or pseudoscalar)-gauge vertices should also be regulated, because consistency would require to have regulated momenta flowing through all vertices, once the propagators entering them contain the regulator. A suitable vertex regularization can be achieved by adding the following off diagonal components to the regulator matrix:
\begin{subequations}
\label{Eq:Roffdiag}
\bea
{\cal R}_{k,A_\alpha \pi}(q)&=&-iZ_{e,k}e (k\hat{q}^\alpha-q^\alpha) \Theta(k^2-q^2) \tilde{\sigma}^a, \\
{\cal R}_{k,A_\alpha \sigma}(q)&=&iZ_{e,k}e (k\hat{q}^\alpha-q^\alpha) \Theta(k^2-q^2) \tilde{\pi}^a.
\eea
\end{subequations}
This is a perfectly legitimate regulator contribution, quadratic in the dynamical variables, as $\tilde{\sigma}^a$ and $\tilde{\pi}^a$ are meant to be nondynamical, homogeneous fields, which are supposed to be set equal to the actual (homogeneous) value of $\sigma^a$ and $\pi^a$, where the effective action is evaluated. In accordance with (\ref{Eq:Roffdiag}), by introducing the notation $q_R^\alpha=\hat{q}^{\alpha}[q+(k-q)\Theta(k^2-q^2)]$, the diagrams take the form of (the first one does not change)
\begin{itemize}
\item $k\tilde{\partial}_k\Big(-\frac{Z_{e,k}^4e^4}{4Z_k^2Z_{A,k}^2}\Big)\int_x (s^as^a+\pi^a\pi^a)^2 \int_q\frac{1}{(q^2+R_k(q))^2}$ \newline
$\times \Big[\delta^{\alpha \beta}-(1-\xi_k)\frac{q^\alpha q^\beta}{q^2}\Big]\Big[\delta^{\beta \alpha}-(1-\xi_k)\frac{q^\beta q^\alpha}{q^2}\Big],$
\item $k\tilde{\partial}_k \frac{Z_{e,k}^4e^4}{2Z_k^2Z_{A,k}^2}\int_x (s^as^a+\pi^a\pi^a)^2 \int_q\frac{1}{(q^2+R_k(q))^3}$ \newline
$\times \Big[\delta^{\alpha \beta}-(1-\xi_k)\frac{q^\alpha q^\beta}{q^2}\Big]\Big[\delta^{\beta \gamma}-(1-\xi_k)\frac{q^\beta q^\gamma}{q^2}\Big]q_R^{\gamma}q_R^{\alpha},$
\item $k\tilde{\partial}_k \Big(-\frac{Z_{e,k}^4e^4}{4Z_k^2Z_{A,k}^2}\Big)\int_x (s^as^a+\pi^a\pi^a)^2 \int_q\frac{1}{(q^2+R_k(q))^4}$ \newline
$\times \Big\{\Big[\delta^{\alpha \beta}-(1-\xi_k)\frac{q^\alpha q^\beta}{q^2}\Big]q_R^{\alpha}q_R^{\beta}\Big\}^2$.
\end{itemize}
Using the equality of the unit vectors, $\hat{q}_R^\alpha=\hat{q}^\alpha$, and performing all differentiations and integrations, the $\xi_k$ dependence indeed cancels, and the sum of the three diagrams contributing to the flow of the effective action becomes
\bea
\label{Eq:e4final}
k\partial_k \Gamma_k &\Leftarrow& \int_x \Big(\frac{\Omega_{\cal D}}{{\cal D}}k^{{\cal D}-4}4({\cal D}-1)Z_k^2 e_k^4\Big)|\Tr(\Phi^\dagger \Phi)|^2.\nonumber\\
\eea
Collecting all contributions from (\ref{Eq:g1ka}), (\ref{Eq:g2ka}), (\ref{Eq:ge2final}), and (\ref{Eq:e4final}), we get
\bea
\label{Eq:g1final}
k\partial_k (Z_k^2g_{1,k})&=&Z_k^2\frac{\Omega_{\cal D}}{{\cal D}}k^{{\cal D}-4}\Big[8(d+4)g_{1,k}^2+16N_Tdg_{1,k}g_{2,k}\nonumber\\
&&\hspace{-0.5cm}+12C_2(A)N_Tg_{2,k}^2-8\xi_ke_k^2g_{1,k}+4({\cal D}-1)e_k^4\Big], \nonumber\\
\\
\label{Eq:g2final}
k\partial_k (Z_k^2g_{2,k})&=&Z_k^2\frac{\Omega_{\cal D}}{{\cal D}}k^{{\cal D}-4}\Big[48g_{1,k}g_{2,k}\nonumber\\
&+&(20N_Td-24C_2(A))g_{2,k}^2-8\xi_ke_k^2g_{2,k}\Big].
\eea
If we set $g_{2,k}\equiv 0$, using the flow of $Z_k$, we get back our earlier results for the $\beta$ functions (see below) for $N$ complex scalar fields with $U(1)$ gauge symmetry ($d=N$) \cite{fejos17}, obtained in the $R_\xi$ gauge. As also discussed in Sec. IVD, in that earlier approach, we used a regulator, which was formulated in terms of eigenmodes and not the original field variables. Now what we see is that in the usual covariant gauge, identical results can only be achieved if the regulator is constructed more like at the level of diagrams. Gauge propagators have to contain a regulator where it was inevitably necessary (to avoid IR divergences), but not if the diagrams at the given order do make sense without it.

That is to say, it is allowed to switch between regularization schemes at different levels of the calculations, as long as the associated regulator functions are legitimate and do not lead to divergences. This could be surprising at first sight, as in the FRG approach, one usually defines the regulator before evaluating any projection of the flow equation. But since any legitimate regulator represents a construction of the scale evolution of the effective action, one may regard switching between regularization schemes as a part of the employed approximation. Since we get the very same results this way compared to the one that associates regulators (in advance) to eigenmodes, we believe that our current approach is justified.

\subsection{$\beta$ functions and fixed points}

Now we analyze the $\beta$ functions of $g_{1,k}$, $g_{2,k}$, and $e_k^2$ and search for fixed points. The $\beta$ functions are defined as the flow of dimensionless couplings, i.e., $\bar{g}_{1,k}=g_{1,k}k^{{\cal D}-4}$, $\bar{g}_{2,k}=g_{2,k}k^{{\cal D}-4}$, $\bar{e}_k^2=e_k^2 k^{{\cal D}-4}$. By definition, they are
\bea
\label{Eq:flowg1}
\beta_{g_1}&\equiv& k\partial_k \bar{g}_{1,k}\nonumber\\
&=&({\cal D}-4)\bar{g}_{1,k}+\frac{k^{{\cal D}-4}}{Z^2_k}k\partial_k(Z_k^2g_{1,k})-2\bar{g}_{1,k}\frac{k\partial_k Z_k}{Z_k}, \nonumber\\  \\
\label{Eq:flowg2}
\beta_{g_2}&\equiv& k\partial_k \bar{g}_{2,k}\nonumber\\
&=&({\cal D}-4)\bar{g}_{2,k}+\frac{k^{{\cal D}-4}}{Z^2_k}k\partial_k(Z_k^2g_{2,k})-2\bar{g}_{2,k}\frac{k\partial_k Z_k}{Z_k}. \nonumber\\ 
\eea
As for the flowing charge, $\bar{e}_k^2=k^{{\cal D}-4}e_k^2$, $e_k^2=e^2Z^2_{e,k}/Z_k^2Z_{A,k}$, we have
\bea
\label{Eq:flowe2}
\beta_{e^2}&\equiv& k\partial_k \bar{e}_k^2\nonumber\\
&=& ({\cal D}-4)\bar{e}_k^2-\bar{e}_k^2\frac{k\partial_k Z_{A,k}}{Z_{A,k}}+\bar{e}_k^2\frac{Z_k^2}{Z_{e,k}^2}k\partial_k\Big(\frac{Z_{e,k}^2}{Z_k^2}\Big) . \nonumber\\
\eea
We assume that the Ward identity $Z_{e,k}=Z_k$ is satisfied, and thus the last term is zero. We will come back to this issue in Sec. IVE, but note that this is a particularly important point, as this last, anomalous term would completely change the nature of $\beta_{e^2}$ had it not been dropped, and due to the fact that ${\cal D}=3$, prevented the existence of fixed points with a finite charge for a small number of scalars. Using  (\ref{Eq:dkZk}), (\ref{Eq:dkZAk}), (\ref{Eq:g1final}), (\ref{Eq:g2final}), we get
\bea
\label{Eq:fullg1}
\beta_{g_1}&=&({\cal D}-4)\bar{g}_{1,k}+\frac{\Omega_{\cal D}}{{\cal D}}\Big\{8(d+4)\bar{g}_{1,k}^2+16N_Td\bar{g}_{1,k}\bar{g}_{2,k}\nonumber\\
&+&12C_2(A)N_T\bar{g}_{2,k}^2+8\bar{e}_k^2\bar{g}_{1,k}\frac{({\cal D}-4)\xi_k-2({\cal D}-1)}{{\cal D}-2} \nonumber\\
&+&4({\cal D}-1)\bar{e}_k^4\Big\}, \\
\label{Eq:fullg2}
\beta_{g_2}&=&({\cal D}-4)\bar{g}_{2,k}+\frac{\Omega_{\cal D}}{{\cal D}}\Big\{(20dN_T-24C_2(A))\bar{g}_{2,k}^2\nonumber\\
&+&48\bar{g}_{1,k}\bar{g}_{2,k}+8\bar{e}_k^2\bar{g}_{2,k}\frac{({\cal D}-4)\xi_k-2({\cal D}-1)}{{\cal D}-2}\Big\}, \\
\label{Eq:fulle2}
\beta_{e^2}&=&({\cal D}-4)\bar{e}_k^2+\frac{\Omega_{\cal D}}{\cal D}\frac{8d}{{\cal D}+2}\bar{e}_k^4.
\eea
We see that in ${\cal D}=4$, the $\xi$ dependence vanishes, but for ${\cal D}\neq 4$, one needs to substitute $({\cal D}-4)\xi_k \rightarrow -2$; see the calculation prior to Eq. (\ref{Eq:xicons}).

There are various fixed points displayed as solutions of the coupled equations of (\ref{Eq:fullg1}), (\ref{Eq:fullg2}), (\ref{Eq:fulle2}). Analytic solutions are available, but we do not list them as they are too long. For chargeless ($\bar{e}_k \equiv 0$) fixed points, there are always two solutions with $\bar{g}_{1,k}\neq 0$, $\bar{g}_{2,k}=0$ (i.e., Gaussian and Wilson-Fisher), and if
\bea
36(C_2(A))^2+12C_2(A)(4-3d)N_T+d^2N_T^2>0,
\eea
then two new ones appear with $\bar{g}_{1,k}\neq 0$, $\bar{g}_{2,k}\neq 0$. For charged fixed points ($\bar{e}_k \neq 0$), in ${\cal D}=3$, there are always two fixed points, where $\bar{g}_{1,k}\neq 0$, $\bar{g}_{2,k}=0$ (these are equivalent to the superconducting and tricritical fixed points found in \cite{fejos16,fejos17}), but we also get two additional ones with $\bar{g}_{2,k}\neq 0$, if
\bea
&&(C_2(A))^2(18000-1440d+36d^2)\nonumber\\
&+&N_TC_2(A)(21600-19920d+3888d^2-36d^3)\nonumber\\
&+&d^2N_T^2(2900-2440d+d^2)>0
\eea
is satisfied. This shows that depending on the structure of the Lie algebra, besides the superconducting transition, models of the form of (\ref{Eq:basicL}) can also show critical behaviors that are characterized by a fixed point for which all couplings are nonzero, $\bar{g}_{1,k}\neq 0$, $\bar{g}_{2,k}\neq 0$, $\bar{e}^2_k\neq 0$.

\begin{figure}
\includegraphics[bb = 150 465 380 570,scale=0.85,angle=0]{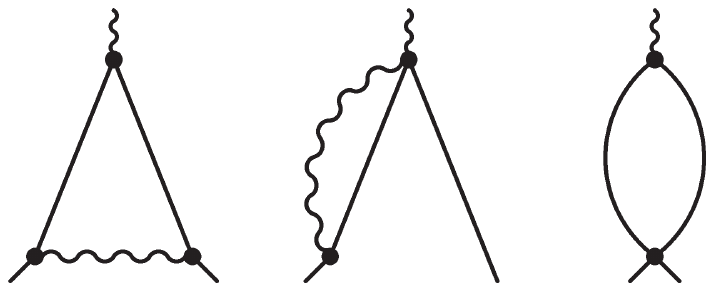}
\caption{Diagrams that contribute to the charge rescaling factor, $Z_{e,k}$. Solid lines refer to the scalar (or pseudoscalar) fields, while the wiggly one corresponds to the gauge field. The external gauge momentum is set to zero for simplicity in the calculation.}
\end{figure}  

\subsection{The $Z_k=Z_{e,k}$ identity}

One of our motivations of using the covariant gauge is that unlike the $R_\xi$ gauge, it does not introduce any asymmetry between the $s^a$ and $\pi^a$ fields. In our earlier study this distinction prevented us from showing in the preferred regularization that the flow equation and the modified Ward-Takahashi identitites (mWTI) lead to the same conclusion regarding the (anomalous) identity between the flow of the charge rescaling factor, $Z_{e,k}$, and the scalar wave function renormalization, $Z_k$. Here, we address this question once again and show that both the flow equation and the (modified) Ward-Takahashi identity leads to the same result.

First, we calculate $k\partial_k Z_{e,k}$, using the flow equation. In Fig. 4, we show which diagrams need to be taken into account to calculate the contribution to the scalar-pseudoscalar-gauge vertex, which leads to the flow of $Z_{e,k}$. The third diagram is identically zero, and the remaining ones yield the following equation:
\bea
&&2p_i k\partial_k Z_{e,k}=Z_{e,k}e_k^2k\tilde{\partial}_k\int_q \frac{1}{(p+q)^2}(q-p)_j(p-q)_m\nonumber\\
&\times&\bigg[\delta_{jm}-(1-\xi_k)\frac{(p+q)_j(p+q)_m}{(p+q)^2}\bigg]\frac{2q_i}{(q^2+R_k(q))^2}\bigg|_{{\cal O}(p)}\nonumber\\
&&\hspace{1.8cm}+Z_{e,k}e_k^2k\tilde{\partial}_k\int_q \frac{4}{(q+p)^2}\frac{(q-p)_j}{q^2+R_k(q)}\nonumber\\
&&\hspace{1.8cm}\times\bigg[\delta_{ij}-(1-\xi_k)\frac{(p+q)_i(p+q)_j}{(p+q)^2}\bigg]\bigg|_{{\cal O}(p)},\nonumber\\
\eea
where the rhs has to be projected onto the ${\cal O}(p)$ piece. We find for the flow of $Z_{e,k}$ the following:
\bea
\label{Eq:dkZek}
k\partial_k Z_{e,k}=\frac{\Omega_{\cal D}}{{\cal D}}8e_k^2Z_{e,k}\bigg[\frac{{\cal D}-1+\xi_k(3-{\cal D})}{{\cal D}-2}+\xi_k\frac{{\cal D}-4}{2{\cal D}}\bigg].\nonumber\\
\eea
If the second term was not present in the bracket, we would get exactly the same flow as of $Z_k$, but this shows that unless we are working in ${\cal D}=4$, $Z_k\neq Z_{e,k}$. Combining (\ref{Eq:dkZek}) with (\ref{Eq:dkZk}), we arrive at
\bea
\label{Eq:ZeZ}
k\partial_k \log \bigg( \frac{Z_{e,k}}{Z_k}\bigg)=\frac{\Omega_{\cal D}}{\cal D}\frac{4({\cal D}-4)}{\cal D}\xi_k\bar{e}_k^2.
\eea

Now we show that Eq. (\ref{Eq:ZeZ}) is compatible with the corresponding regulator modified Ward-Takahashi identity. In the presence of the ${\cal R}_k$ function, the master equation generating these identities for a symmetry transformation denoted by $\delta$ is (see derivation in \cite{fejos17})
\bea
\label{Eq:WTI}
\langle\delta S[\hat{\varphi}]\rangle-\int\langle \delta \hat{\varphi}\rangle\frac{\delta \Gamma_k}{\delta \varphi}&=&-\langle \delta \int \hat{\varphi}^\dagger {\cal R}_k \hat{\varphi}\rangle \nonumber\\
&&+\int \langle \delta \hat{\varphi}\rangle \frac{\delta}{\delta \varphi}\int \varphi^\dagger {\cal R}_k \varphi,
\eea
where $\varphi$ denotes the collection of fields of a field theory, and $\delta S$ is a term that explicitly breaks the symmetry in question in the classical action (e.g., the gauge fixing term). The fluctuating fields are denoted by hats, and $\langle ... \rangle$ refers to averaging. When one projects both sides of (\ref{Eq:WTI}) onto various operators, identities between $n$-point functions can be found. Considering our current theory, denoting the gauge transformation parameter by $\theta$, the projection onto $\sim\! \Tr[\Phi^\dagger \Phi]$ leads the rhs to
\bea
\label{Eq:Wardrhs}
\rhs&=&-ie \int_p \theta(-p)\int_q \Tr[\Phi^\dagger(q) \Phi(q+p)](p^2+2p\cdot q) \nonumber\\
&\times&\int_l \frac{R_k(l)}{l^2(l^2+R_k(l))^2}Z_k\xi_ke_k^2\frac{2({\cal D}-2)}{{\cal D}},
\eea
while for the lhs, we get
\bea
\lhs&=&-ie\int_p \theta(-p)\int_q \Tr[\Phi^\dagger(q) \Phi(q+p)](p^2+2p\cdot q) \nonumber\\
&\times&(Z_{e,k}-Z_k).
\eea
These expressions are obtained by analogous calculations as of \cite{fejos17}, and we do not go into the details. Since $\int_l R_k(l)/l^2=2\Omega_{\cal D}k^{\cal D}/{\cal D}({\cal D}-2)$, we get 
\bea
\label{Eq:Zratioward}
\frac{Z_{e,k}}{Z_k}=1+\frac{\Omega_{\cal D}}{{\cal D}}\frac{4}{{\cal D}}k^{{\cal D}-4}\xi_k e_k^2,
\eea
which leads to
\bea
\label{Eq:Zratioflow}
k\partial_k \log \bigg( \frac{Z_{e,k}}{Z_k}\bigg)=\frac{\Omega_{\cal D}}{\cal D}\frac{4({\cal D}-4)}{\cal D}\xi_k\bar{e}_k^2+{\cal O}(e^4),
\eea
being equivalent to (\ref{Eq:ZeZ}) at leading order. Note, however, that, (\ref{Eq:Zratioward}) also specifies the initial condition at the UV scale $k=\Lambda$, while (\ref{Eq:Zratioflow}) only describes the scale evolution of $Z_{e,k}/Z_k$.

Equations (\ref{Eq:Zratioward}) and (\ref{Eq:Zratioflow}) show that the Ward-Takahashi identity, $Z_{e,k}=Z_k$, can only be maintained for $\xi_k\equiv 0$, but that is not a legitimate choice as it leads to a discrepancy related to the flow of the gauge wave function renormalization in ${\cal D}\neq 4$; see once again the incompatibility between (\ref{Eq:diagexp}) and (\ref{Eq:diagsum}). The only way one can circumvent the problem of anomalous contributions proportional to  $Z_{e,k}/Z_k$, see, e.g., the flow of the charge (\ref{Eq:flowe2}), is to completely discard the flow of $Z_{e,k}$, and impose the identity $Z_{e,k}=Z_k$ at all scales. As already stressed, this is important, because the anomaly related to $Z_{e,k}\neq Z_k$ can make IR fixed points disappear; see again (\ref{Eq:flowe2}) [and its combination with  (\ref{Eq:flowg1}), (\ref{Eq:flowg2})].

\section{Conclusions}

In this paper, we have analyzed the renormalization group flows of Abelian gauge theories with multicomponent scalars $(\Phi)$ that are built up from elements of Lie algebras. We have investigated if the local potential approximation (LPA) is a one-loop closed truncation of the RG flows, for classes of theories, where two quartic couplings, belonging to operators $\sim\!|\Tr (\Phi^\dagger \Phi)|^2$ and $\sim\!\! \Tr(\Phi^\dagger \Phi \Phi^\dagger \Phi)$, are present. We have found that, in principle, it is not true that the flow equation preserves the structure of the classical Lagrangian; hence the LPA is not generically one-loop closed. We have also found, however, that if a simple, compact Lie algebra is extended by a $U(1)$ factor, then one never encounters the aforementioned problem. The key ingredient was that for such systems, not only the commutator, but also the anticommutator of two algebra elements belong to the algebra itself.

Gauge effects have been explored in the standard covariant gauge, and our earlier results in the $R_\xi$ gauge for the $\beta$ functions of $N$ complex scalars have been recovered; thus, we have established connections between two different approaches. This was nontrivial in the sense that it has turned out that the preferred regularization procedure in the $R_\xi$ gauge (where regulators were attached to eigenmodes \cite{fejos17}) corresponds to a scheme in the covariant gauge in which regularizing propagators depends on the actual diagram in question. We have found that no regulator needs to be associated to the gauge field, except for the ${\cal O}(e^4)$ contributions in the self coupling flow. In these terms, momenta both in gauge and scalar lines need to be regulated. This argues that it is not unnatural to switch between regularization schemes at different levels of the calculations, and this can lead to reliable results.

Finally, we have showed that, in the covariant gauge, the anomalous breaking of the Ward-Takahashi identity between the scalar wave function renormalization and the charge rescaling factor leads to the same result as the flow equation. Note that the Ward-Takahashi identity also specifies initial conditions at the UV scale, while the flow equation only describes their scale evolution.

It would be interesting to generalize our method to non-Abelian gauge theories. As a natural continuation, one could analyze the Ginzburg-Landau theory of color superconductivity, which is in essence an analogous model to that of the current study, but with $SU(3)$ gauge symmetry. A final goal would be to investigate fermionic effects, and see how these analyses can be carried out in QCD itself.

\section*{Acknowledgements}

This work was supported by the Ministry of Education, Culture, Sports, Science (MEXT)-Supported Program for the Strategic Research Foundation at Private Universities “Topological Science” (Grant No. S1511006). G.F. was also supported by the Hungarian National Research, Development and Innovation Office (Project No. 127982), and by the János Bolyai Research Scholarship of the Hungarian Academy of Sciences. T.H. was partially supported by JSPS Grand-in-Aid for Scienfitic Research (S), Grant No. 18H05236.

\makeatletter
\@addtoreset{equation}{section}
\makeatother 

\renewcommand{\theequation}{A\arabic{equation}} 

\appendix
\section{Lie algebras}

\hspace{-0.5cm} {\bf Case of simple Lie algebras}\newline \newline
Here, we discuss some properties of finite $(d)$ dimensional simple Lie algebras that generate compact Lie groups. In the latter half of this appendix, we also touch upon these algebras with one additional $U(1)$ factor. According to Ado's theorem, the generators, $T_i$, which span the algebra, are necessarily matrices, and the Lie bracket, $[\hspace{0.1cm},\hspace{0.1cm}]$ is the commutator. The anticommutator will be denoted by $\{\hspace{0.1cm},\hspace{0.1cm}\}$. In principle, the product of two generators spills out of the algebra, but in the fundamental representation, one has
\bea
\label{Eq:appTT}
T_iT_j = \frac{N_T}{4}\delta_{ij}{\bf 1}+\frac12(d_{ijk}+if_{ijk})T_k,
\eea
where ${\bf 1}$ is the unit matrix, $N_T=2/\Tr({\bf 1})$, and the $d_{ijk}$ and $f_{ijk}$ are totally symmetric and antisymmetric structure constants, respectively. Here, we used compactness and thus $(T_i)^\dagger=T_i$, and furthermore, $T_i$ are normalized as $\Tr(T_iT_j)=\delta_{ij}/2$. We also assumed that the algebra is simple; thus, $T_i$ are traceless. Since matrix multiplication is associative, for any $X$, $Y$, $Z$ elements of the algebra,
\bea
[X,YZ]+\{Y,ZX\}-\{Z,XY\}=0,
\eea
from which one derives
\bea
[X,YZ]+[Z,XY]+[Y,ZX]=0,
\eea
and also the Jacobian identity (as it should be),
\bea
\label{Eq:Jacob}
[X,[Y,Z]]+[Z,[X,Y]]+[Y,[Z,X]]=0.
\eea
Using these identities, one can also derive more,
\bea
\label{Eq:id1}
[\{X,Y\},Z]+[\{Y,Z\},X]+[\{Z,X\},Y]=0,
\eea
\bea
\label{Eq:id2}
[Z,[X,Y]]-\{Y,\{Z,X\}\}+\{X,\{Y,Z\}\}=0.
\eea
From (\ref{Eq:Jacob}), (\ref{Eq:id1}), and (\ref{Eq:id2}), one gets the following restrictions for the structure constants (use $X=T_i$, $Y=T_j$, $Z=T_k$!):
\begin{subequations}
\label{Eq:ids}
\bea
0&=&f_{ilm}f_{mjk}+f_{jlm}f_{imk}+f_{klm}f_{ijm}, \\
0&=&f_{ilm}d_{mjk}+f_{jlm}d_{imk}+f_{klm}d_{ijm}, \\
f_{ijm}f_{klm}&=&d_{ikm}d_{jlm}-d_{jkm}d_{ilm}+N_T(\delta_{ik}\delta_{jl}-\delta_{jk}\delta_{il}). \nonumber\\
\eea
\end{subequations}
Equations (\ref{Eq:ids}) are the starting point for deriving several useful identities. We denote by $C_2(A)$, the value of the Casimir operator in the adjoint representation, i.e., $f_{ijk}f_{ljk}=C_2(A)\delta_{il}$. Without going into details, using Eqs. (\ref{Eq:ids}), the following equations can be obtained:
\newline
\begin{subequations}
\label{Eq:ids2}
\bea
d_{ijm}d_{kjm}&=&[N_T(d-1)-C_2(A)]\delta_{ik}, \\
f_{lni}f_{ikm}f_{mjl}&=&-\frac{C_2(A)}{2}f_{nkj}, \\
d_{mik}d_{knj}f_{jlm}&=&\frac{N_T(d-1)-C_2(A)}{2}f_{inl}, \\
f_{inl}f_{ljm}d_{mki}&=&-\frac{C_2(A)}{2}d_{njk}, \\
d_{ikl}d_{lnm}d_{mji}&=&[N_T(d-2)-\frac32C_2(A)]d_{knj}, 
\eea
\end{subequations}
These identities are fairly simple to derive. We also need fourfold sums of $d_{ijk}$, which, in turn, are very tedious to obtain. First, one derives an expression for the totally symmetric combination,
\bea
\label{Eq:id3}
d_{i(aj}d_{jbk}d_{kcl}d_{ld)i}&=&\big[(d-3)N_T-\frac74C_2(A)\big] d_{m(ab}d_{cd)m} \nonumber\\
&+&N_T\big[(d-1)N_T-C_2(A)\big]\delta_{(ab}\delta_{cd)}, \nonumber\\
\eea
and then gets
\bea
\label{Eq:id4}
d_{iaj}d_{jbk}d_{kcl}d_{ldi}&=&\Big[\frac12(d-3)N_T-\frac34C_2(A)\Big]\nonumber\\
&\times&(d_{abm}d_{mcd}+d_{adm}d_{mbc}) \nonumber\\
&-&\frac{C_2(A)}{4}d_{acm}d_{mbd} \nonumber\\
&+&\frac{N_T}{2}[(d-1)N_T-C_2(A)] \nonumber\\
&\times&(\delta_{ab}\delta_{cd}+\delta_{ad}\delta_{bc}).
\eea
These formulas are generalizations of those found in \cite{azcarraga98}.

\hspace{-0.35cm}{\bf Case of simple Lie algebras with one $U(1)$ factor}
\newline \newline
Now we turn our attention to simple algebras extended with one $U(1)$ factor. The corresponding generator is denoted by $T_0=\sqrt{N_T}/2\cdot {\bf 1}$, and the new structure constants are $d_{ij0}=\sqrt{N_T}\delta_{ij}$ and $f_{ij0}=0$. Equation (\ref{Eq:appTT}) changes to
\bea
\label{Eq:appTT2}
T_iT_j = \frac12(d_{ijk}+if_{ijk})T_k,
\eea
and Eqs. (\ref{Eq:ids}) become
\begin{subequations}
\bea
0&=&f_{ilm}f_{mjk}+f_{jlm}f_{imk}+f_{klm}f_{ijm}, \\
0&=&f_{ilm}d_{mjk}+f_{jlm}d_{imk}+f_{klm}d_{ijm}, \\
0&=&f_{ijm}f_{klm}-d_{ikm}d_{jlm}+d_{jlm}d_{ilm}
\eea
\end{subequations}
[only the last one changed compared to (\ref{Eq:ids})]. Note that now the sums also run over the index $m=0$, and we have the modified identity, $f_{ijk}f_{ljk}=C_2(A)\delta_{il}(1-\delta_{i0}\delta_{l0})$. As a result, Eqs. (\ref{Eq:ids2}), (\ref{Eq:id3}), and (\ref{Eq:id4}) do not maintain their forms. We get the following new identities:
\begin{subequations}
\bea
d_{ijm}d_{kjm}&=&\big(N_Td-C_2(A)(1-\delta_{i0}\delta_{k0})\big)\delta_{ik}, \\
f_{lni}f_{ikm}f_{mjl}&=&-\frac{C_2(A)}{2}f_{nkj}, \\
d_{mik}d_{knj}f_{jlm}&=&\frac{N_Td-C_2(A)}{2}f_{inl}, \\
f_{inl}f_{ljm}d_{mki}&=&-\frac{C_2(A)}{2}d_{njk}+\frac12C_2(A)\sqrt{N_T} \nonumber\\
&\times&(\delta_{n0}\delta_{jk}+\delta_{j0}\delta_{nk}-\delta_{k0}\delta_{nj}),\\
d_{ikl}d_{lnm}d_{mji}&=&(N_Td-\frac32C_2(A))d_{njk}+\frac12C_2(A)\sqrt{N_T} \nonumber\\
&\times&(\delta_{n0}\delta_{jk}+\delta_{j0}\delta_{nk}-\delta_{k0}\delta_{nj}).
\eea
\end{subequations}
As for the fourfold sum,
\bea
d_{iaj}d_{jbk}d_{kcl}d_{ldi}&=&\big(N_Td-\frac32C_2(A)\big)d_{abm}d_{mcd}\nonumber\\
&&\hspace{-1cm}+\big(\frac12N_Td-C_2(A)\big)[d_{adm}d_{mbc}+d_{acm}d_{mbd}]\nonumber\\
&&\hspace{-1cm}+\frac{C_2(A)N_T}{2}(\delta_{ab}\delta_{cd}+\delta_{ac}\delta_{bd}+\delta_{ad}\delta_{bc}) \nonumber\\
&&\hspace{-1cm}+\frac12C_2(A)\sqrt{N_T}(\delta_{a0}d_{bcd}+\delta_{b0}d_{acd} \nonumber\\
&&\hspace{+1.2cm}+\delta_{c0}d_{abd}+\delta_{d0}d_{abc}),
\eea
which can be derived as explained above.

\renewcommand{\theequation}{B\arabic{equation}} 
\section{Calculation of matrix elements}

In this appendix we list all the elements of the matrices $A=[(\Tr(\Phi^\dagger \Phi))^{2}]''$, $B=[\Tr(\Phi^\dagger \Phi \Phi^\dagger \Phi)]''$. We start with the case of simple algebras, for which these matrices were introduced in Sec. IIIA. One differentiates Eqs. (\ref{Eq:inva}) with respect to the fields $s^i$ and $\pi^j$ to obtain the elements of the $A$ and $B$ matrices. As for $A$, one has
\begin{subequations}
\label{Eq:Aderiv}
\bea
(A_{ss})_{ij}&=&(s^as^a+\pi^a\pi^a)\delta^{ij}+2s^is^j, \\
(A_{\pi\pi})_{ij}&=&(s^as^a+\pi^a\pi^a)\delta^{ij}+2\pi^i\pi^j, \\
(A_{s\pi})_{ij}&=&2s^i\pi^j,
\eea
\end{subequations} 
while for $B$, one arrives at
\begin{subequations}
\label{Eq:Bderiv}
\bea
(B_{ss})_{ij}&=&\frac12s^as^bD_{abij}+\frac12\pi^c\pi^d[4\tilde{D}_{ij,cd}-D_{ijcd}], \nonumber\\ \\
(B_{\pi\pi})_{ij}&=&\frac12\pi^a\pi^bD_{abij}+\frac12s^cs^d[4\tilde{D}_{ij,cd}-D_{ijcd}], \nonumber\\ \\
(B_{s\pi})_{ij}&=&s^a\pi^b[4\tilde{D}_{ai,bj}-D_{aibj}].
\eea
\end{subequations} 
For the case of simple algebras extended with one $U(1)$ factor, introduced in Sec. IIID, the $B$ matrix elements modify as follows:
\begin{subequations}
\label{Eq:Bderiv2}
\bea
(B_{ss})_{ij}&=&\frac12s^as^bD_{abij}+\frac12\pi^c\pi^d[4\tilde{D}_{ij,cd}-D_{ijcd}], \nonumber\\ 
&+&N_T\Big[\frac12(s^as^a+\pi^a\pi^a)\delta^{ij}+s^is^j \nonumber\\
&&\hspace{0.7cm}+\pi^a\pi^a\delta^{ij}-\pi^i\pi^j\Big], \\
(B_{\pi\pi})_{ij}&=&\frac12\pi^a\pi^bD_{abij}+\frac12s^cs^d[4\tilde{D}_{ij,cd}-D_{ijcd}], \nonumber\\
&+&N_T\Big[\frac12(s^as^a+\pi^a\pi^a)\delta^{ij}+\pi^i\pi^j \nonumber\\
&&\hspace{0.7cm}+s^as^a\delta^{ij}-s^is^j\Big], \\
(B_{s\pi})_{ij}&=&s^a\pi^b[4\tilde{D}_{ai,bj}-D_{aibj}].
\eea
\end{subequations} 
With the help of (\ref{Eq:Aderiv}), (\ref{Eq:Bderiv}), and (\ref{Eq:Bderiv2}), one is set to calculate $\Tr(V_k''V_k'')$ via (\ref{Eq:TrVVform}).

Finally, for the sake of completeness, we list the elements of the $C$ matrix [only relevant for the $SU(3)$ case], defined in Sec. IIIC. The definition of the $C$ matrix is, see also (\ref{Eq:defC}),
\bea
C = (s^as^a\pi^b\pi^b-(s^a\pi^a)^2)''.
\eea
The matrix elements are
\begin{subequations}
\bea
(C_{ss})_{ij}&=&2(\pi^a\pi^a\delta_{ij}-\pi^i\pi^j), \\
(C_{\pi\pi})_{ij}&=&2(s^as^a\delta_{ij}-s^is^j), \\
(C_{s\pi})_{ij}&=&4s^i\pi^j-2s^j\pi^i-2s^a\pi^a\delta_{ij}.
\eea
\end{subequations}

\end{document}